\documentclass[preprint,12pt]{elsarticle}
\usepackage{hyperref}
\usepackage{times}
\usepackage{amscd,amsmath,amssymb,amsthm}
\usepackage{color} 

\usepackage{mathrsfs}
\usepackage{graphicx}

\usepackage{helvet}
\usepackage{courier}

\usepackage{algorithm}
\usepackage[noend]{algorithmic}
\usepackage{verbatim}
\usepackage{flafter}

\begin{document}

\journal{arXiv}

\begin{frontmatter}


\title{A verifiable multi-party quantum key distribution protocol based on repetitive codes}

\author[1]{Lei Li\corref{cor1}}
\ead{lilei2018@stu.xidian.edu.cn}
\author[2]{Zhi Li}
\ead{zhli@xidian.edu.cn}
\address[1]{School of Mechano-Electronic Engineering, Xidian University, Xi'an, 710071, China}

\cortext[cor1]{Corresponding Author}
\begin{abstract}
A multi-party quantum key distribution protocol based on repetitive code is designed for the first time in this paper. First we establish a classical (t, n) threshold protocol which can authenticate the identity of the participants, and encode the classical key sequence in accordance with this repetitive code. Then unitary transformation of the quantum state sequence corresponding to this encoded sequence is carried out by using the parameters from this (t, n) threshold protocol. Furthermore, we derive two thresholds for whether or not reserving the measured values of the received sequence, and extract the classical subkey sequence from the measured values conforming to these two threshold conditions. This protocol can authenticate the identity of the participant, resist the attack from the internal and external participants, and do not need the decoy state particles when testing the eavesdropper, which is more efficient than the similar protocols, and also saves the quantum resources.
\end{abstract}

\end{frontmatter}

\section{Introduction}
Quantum Key Distribution (QKD) is a technique that permits two parties, who share no secret information initially, to communicate over an open channel and to establish between themselves a shared secret sequence of bits. Since C. H. Bennett and G. Brassard first proposed the complete QKD protocol in 1984, QKD has experienced the improvement of theoretical assumptions and schemes. At the same time, due to the imperfect light source and measuring equipment in the actual quantum key distribution protocol, QKD has also experienced many improvements in practical applications \cite{A3,A5,A6,A7,A8,A9,A10,A11,A12,A13,A14}. In 2018, \cite{A15} proposed a phase-matched quantum key distribution scheme which can transcend the limit of the linear key rate, which is not only guaranteed in terms of security and practicality, but also significantly improved in terms of transmission distance. So it is a high-performance quantum key distribution protocol. These results show that the theory and experimental technology for point-to-point QKD are becoming mature.

With the maturity of quantum key distribution technology between two parties, people begin to pay close attention to the expansion of the quantum key distribution protocol, namely research multiparty quantum key distribution (MQKD) protocol \cite{A11,A12,A13} and multiparty quantum key agreement protocol (MQKA) \cite{A14,A15,A16,A17,A18,A19}. MQKD protocol is composed of one party distributing keys to the other parties, and each participant in MQKA agreement made the same contribution to the formation of the shared secret. However, with the increasing number of participants, MQKA protocol and MQKD protocol are faced with two common problems: one is how to ensure the security of information, the other is the efficiency of the protocol.

With regard to the efficiency of the protocol, the paper \cite{A19} uses the cluster state of four qubits as the quantum resource and performs X operation to generate the shared key. Thus, compared with other multi-party QKA protocols \cite{A15,A16}, the protocol from \cite{A19} is more efficient. However, it uses more quantum resources.

With respect to ensuring the security of information, we know that there are three factors for warranting the security of information: in order to assure that the information in the transmission process is not eavesdropped by others, transmission encryption should be carried out; In order to ensure that the identity of the authorized participant is not stolen by others, encryption algorithm can be used for identity demonstration; To make sure that the transmitted content is not tampered with, encryption algorithms can be used for digital authentication. These three elements must also be ensured in the relevant protocols of multi-party quantum keys. At present, most of the protocols can encrypt information or prevent eavesdropping during transmission, but the latter two aspects are difficult to satisfy simultaneously. However, in real life, there may be dishonest participants. Therefore, in order to obtain a truly secure key, it is necessary to verify the authenticity of the identity of the participant and the quantum key obtained. Literature \cite{A20} studied quantum key verification for the first time, and then various verifiable quantum key distribution protocols was proposed \cite{A21,A22,A23}.

The above protocols only realize key authentication, but, in realistic cases, the authentication problem also needs to be solved urgently. In 2014, Guan et al. \cite{A24} proposed a three-party verifiable quantum key distribution protocol based on single photon, and implemented authentication and key authentication on the star network topology. However, this protocol lacks extensibility. Therefore, how to design a multi-party quantum key distribution protocol that can ensure information security, further improve the efficiency, and save quantum resource, is also a problem that is worth studying.

In this paper, we propose a new verifiable MQKD protocol based on unbiased basis. The agreement is composed of classical network and quantum network, in which classical networks adopt Shamir threshold scheme based on binary polynomial. In the quantum network part of this protocol, the secret recovered by the participant and the parameters related to the session keys are derived from this classical protocol to ensure that the designed quantum key distribution protocol has the verifiability of the identity of the participant and the security of the information. Several contributions to this agreement are as follows:

(1) The protocol has high efficiency. For the first time, the repeated code theory is used to encode the quantum information sequence, so that the eavesdropping can be prevented without the need of deceptions in the protocol. Therefore, the protocol has high efficiency and saves quantum resources.

(2) The protocol has identity authentication function. The identity authentication between participants is guaranteed by using the relevant data in binary polynomial theory as the relevant parameters in the quantum key distribution protocol.

(3) The protocol has the ability to verify the information. The private key $s$ generated in the Shamir threshold scheme is used as the secret key of the hash function to verify the accuracy of the information in the quantum key distribution protocol.

(4) The protocol has scalability. In the quantum system model, we assume that each terminal is capable of generating, manipulating, and measuring a single photon. Under such conditions, our agreement is suitable for multi-party participants.

A comparison is made between several current multi-party QKA protocols \cite{A25,A26,A27,A28}. Due to the use of the repetitive code technology and the classical binary polynomial theory, our scheme has significantly improved efficiency, less complexity, and ensures security and identity authentication between participants.

The rest of this paper is organized as followed. In Sec.2, we introduce the relevant conclusions about the unbiased basis. In Sec.3 an improved secret sharing protection model is discussed. In Sect.4 we give a detailed description of the quantum key distribution protocol. An example of this protocol is given in Sec.5. Security analysis of the protocol is discussed in Sec.6. Finally, in Sec.7 we give a short conclusion.

\section{MUBs and their related properties}
Mutually unbiased bases (MUBs) is an important tool in many quantum information
processing. There have been some results on MUBs \cite{B29,B30}.

We define two bases $A_0=\{|\varphi^0_1\rangle,|\varphi^0_2\rangle,\cdots,|\varphi^0_d\rangle\}$ and $A_1=\{|\varphi^1_1\rangle,|\varphi^1_2\rangle,\cdots,$ $|\varphi^1_d\rangle\}$ over a $d$-dimensional complex space to be mutually unbiased if the inner products between all possible vector pairs have the same magnitude:

\begin{equation}
|\langle\varphi^0_l|\psi^1_j\rangle|=\frac{1}{\sqrt{d}},
\label{Eq:3}
\end{equation}

\begin{flushleft}
where $l,j=1,2,\cdots,d$.
\end{flushleft}
\textbf{Definition 1} A set of orthonormal bases $B=\{B_0,B_1,\cdots,B_m\}$ is said to be a set of MUBs if the elements of $B$ are non-biased relative to each other.

Wootters et al. \cite{B29} pointed out that the maximum number of MUBs in $p$-dimensional complex space is $p+1$. It has been pointed out \cite{B29,B30} that there are at most $p+1$ groups of mutually unbiased bases if the dimension of a quantum system is an odd prime number $p$, and one of these bases is a computational basis, i.e.,

\begin{equation}
\{|l\rangle|l\in D\}, \mbox{where } D=\{0,1,\cdots,p-1\}.
\label{Eq:2}
\end{equation}

\begin{flushleft}
The remaining $p$ group basis can be expressed as:
\end{flushleft}

\begin{equation}
|v^{(j)}_l\rangle=\frac{1}{\sqrt{p}}\sum^{p-1}_{k=0}\omega^{k(l+jk)}|k\rangle,
\label{Eq:3}
\end{equation}

\begin{flushleft}
where $\omega=e^{2\pi i/p}$,$l,j\in D$, and $|v^{(j)}_l\rangle$ represents the $l$-th vector in the $j$-th group bases.
\end{flushleft}

These $p+1$ group of unbiased bases meet the following conditions:

\begin{eqnarray*}
\langle v^{(j)}_l|v^{(j')}_l\rangle=\frac{1}{\sqrt{p}},j\neq j'.
\end{eqnarray*}

In Ref.\cite{B29}, the encoding operation consists of two unitary operators, $X$ and $Y$, which are depicted as follows:
\begin{eqnarray*}
X=\sum^{p-1}_{m=0}\omega^m|m\rangle \langle m|, Y=\sum^{p-1}_{m=0}\omega^{m^2}|m\rangle \langle m|.
\end{eqnarray*}

\begin{flushleft}
Using Eq.(1) and Eq.(3), we can get
\end{flushleft}

\begin{equation}
\begin{aligned}
X^xY^y|v^{(j)}_l\rangle &=X^x(\sum^{p-1}_{m=0}\omega^{ym^2}|m\rangle \langle m|)(\frac{1}{\sqrt{p}}\sum^{p-1}_{k=0}\omega^{k(l+jk)}|k\rangle)\\
&=\frac{1}{\sqrt{p}}\sum^{p-1}_{m=0}\omega^{xm+ym^2}|m\rangle\langle m|\sum^{p-1}_{k=0}\omega^{k(l+jk)}|k\rangle\\
&=\frac{1}{\sqrt{p}}\sum^{p-1}_{k=0}\omega^{k((l+x)+(j+y)k)}|k\rangle\\
&=|v^{(j+y)\mbox{mod p}}_{(l+x)\mbox{mod p}}\rangle.
\end{aligned}
\label{Eq:4}
\end{equation}

For the convenience, the operator $X^x Y^y$ is written as $U^y_x$, i.e., $U^y_x|v^{(j)}_l\rangle=|v^{(j+y)}_{(l+x)}\rangle$.

\section{Improved secret-sharing protection model}

In 2017, Lein et al.\cite{C32} proposed the protected secret sharing (PSS) scheme, in which participants include a trusted distributor Alice, $n$ share holders Bob$_1,\cdots$,Bob$_n$ and some internal or external adversaries. The received shares by the shareholder may be used for two purposes:

(a) reconstruct the original secret;

(b) establish the pairwise session keys among shareholders, which are used to establish a secure channel between each pair of shareholders in order to exchange the sub-shares during secret reconstruction.

\begin{flushleft}
This scheme uses binary polynomials to realize the mutual authentication between each pair of shareholders with high efficiency.
\end{flushleft}

However, the condition that the proportion $\varepsilon>{2/3}$ does not guarantee the correctness of this scheme in our opinion, that is to say, this condition cannot ensure that secret can always be reconstructed, and any internal adversary using the false sub-shares in the secret reconstruction phase can be identified. Because the scheme in \cite{C32} is a $(t, n)$ threshold secret sharing one, which means that $t$ or more participants show their sub-shares when the secret is needed to be recovered. But all participants in\cite{C32} were required to show their sub-shares in the recovery secret stage in order to identify dishonest participants. Specifically, while $t$ participants were restoring the secret, if there were $t-1$ dishonest participants, the $t$-th honest participant had to borrow these sub-shares from $n-t+1$ honest participants to restore the secret. However, this is incompatible with the definition of $(t, n)$ threshold secret sharing scheme. So we first have to make some improvements of the scheme in \cite{C32}.

In the new improved scheme, the secret and the session key between participants will be used for the parameters in our quantum key distribution protocol. The improved scheme is designed by using the following asymmetric bivariate polynomials $F(x,y)$, where the degree of $F(x,y)$ on $x$ is at most $t-1$, and that on $y$ is at most $h-1$. $F(x,y)$ can be expressed as
\begin{equation}
F(x,y)=a_{0,0}+a_{1,0}x+a_{0,1}y+\cdots+a_{t-1,h-1}x^{t-1}y^{h-1},
\label{Eq:5}
\end{equation}

\begin{flushleft}
where $a_{i,j}\in F_p$,$\forall i,j\in [0,1,\cdots,t-1]$.
\end{flushleft}

In this scheme, $s=F(0,0)$ is the secret to be shared by the participants, where $0<s<p$. Distributor Alice computes a pair of sub-shares $F(x,x_i)$ and $F(x_i,y)$ for each shareholder Bob$_i$, and $x_i$ is the public information of shareholder Bob$_i$, $i=0,1,\cdots,n$, where $x_0$ is the public identity of Alice (or Alice can also be called Bob$_0$ for ease of description). Distributor Alice sends $s$ pair of sub-shares $\{F(x,x_i),F(x_i,y)\}$ to each shareholder Bob$_i$ over the secure channel. It is important to emphasize that this secure channel must ensure that there is no leakage of the pair sub-shares, so that this secure channel can be implemented through direct quantum communication.

\subsection{The improved Model}
For narrative convenience, suppose that distributor Alice wants $t$ participants Bob$_1$,Bob$_2\cdots$, Bob$_t$ from set A=\{Bob$_1$,Bob$_2\cdots$, Bob$_n$\} to recover the private key $s=F(0,0)$.

\textbf{Step 1} Calculates two shared keys for each pair of share holders. For example, shareholder Bob$_i$ can compute $F(x_i,x_j)$ and $F(x_j,x_i)$ from his paired subshares $\{F(x,x_i),F(x_i,y)\}$; similarly, $Bob_j$ can compute $F(x_i,x_j)$ and  $F(x_j,x_i)$. Thus Bob$_i$ and Bob$_j$ can have a pair of shared keys $F(x_i,x_j)$ and  $F(x_j,x_i)$, where $i<j$.

\textbf{Step 2} Each shareholder Bob$_i$ calculates the Lagrangian component $\delta_i$ using its sub-share $F(x_i,y)$, where
\begin{eqnarray*}
\delta_i=F(x_i,0)\prod^t_{j=1,j\neq i}\frac{-x_j}{x_i-x_j}\mbox{mod }p.
\end{eqnarray*}

\textbf{Step 3} For each pair of share holders, they construct a secure channel using a shared key, and then use this channel to exchange Lagrange shares. For example, shareholder Bob$_i$ calculates $c_{i,j}=E_{F(x_i,x_j)}(\delta_i)$, where $E_{F(x_i,x_j)}(\delta_i)$ means encrypted using the one-time pad about $\delta_i$ with secret $F(x_i,x_j)$ and sends $c_{i,j}$ to shareholder Bob$_j$ over the authenticated broadcast channel C. Similarly, Bob$_j$ uses the shared key $F(x_j,x_i)$ to encrypt her sub-share $\delta_j$ through a one-time pad, and uses the authenticated channel C to send $c_{j,i}$ to the share holder by Bob$_i$.

\textbf{Step 4} The shareholder Bob$_i$ receives the cryptogram $c_{i,j}$, where $j\in\{1,2,\cdots,$ $t\}\backslash\{i\}$. $D_{F(x_j,x_i)}(c_{j,i})=\delta_i$ can be decrypted separately, where $D_{F(x_j,x_i)}(c_{j,i})$ represents using the key $F(x_j,x_i)$ to decrypt $c_{j,i}$.

\textbf{Step 5} The shareholder Bob$_i$ sends $H_{F(x_j,x_i)}(x_i,\delta_j)$ to Alice, and Alice tells Bob$_i$ after verifying whether this message is correct, $j\in\{1,2,\cdots,t\}\backslash\{i\}$.

\textbf{Step 6} If $\delta_j$ is correct, $j\in\{1,2,\cdots,t\}\backslash\{i\}$, each shareholder Bob$_i$ computes the secret $s=\sum^t_{j=1}\delta_j$.

Thus, the sub-share received by shareholder Bob$_i$ can then follow the steps described above to achieve proposes $(a)$ and $(b)$.

\subsection{Security analysis}
\textbf{Theorem 1} (Correctness) The proposed scheme achieves the correctness property. That is, the correct private keys $s$ can always be reconstructed and any participant who uses false sub-shares in the sub-share reconstruction phase can be identified.

\textbf{Proof} In the classical protocol section, suppose that participants Bob$_1,\cdots$, Bob$_t$ want to recover the private key $s$. According to Step 4 of Section 2.2 in this paper, shareholder Bob$_i$ gets $\delta_j$ from Bob$_j$, $j\in\{1,2,\cdots,t\}\backslash\{i\}$, and Bob$_i$ will verify the authenticity of $\delta_j$ to Alice in Step 5. When Bob$_i$ gets the real $\delta_j$, Bob$_i$ can get the private key $s=\sum^t_{i=1}\delta_i$ in Step 6. This is because $F(x,0)$ is a univariate polynomial with the highest power $t-1$, and it is known by the Lagrange interpolation formula that,
\begin{eqnarray*}
s=F(0,0)=\sum^t_{i=1}F(x_i,0) \prod^t_{j=1,j\neq i}\frac{-x_j}{x_i-x_j}(\mbox{mod}p)=\sum^t_{i=1}\delta_i(\mbox{mod}p).
\end{eqnarray*}

Thus, the private key $s$ can always be recovered accurately, and any participant using a false sub-share during the sub-share reconstruction phase can be identified by the hash function $H_{F(x_i,x_0)}$ between participant Bob$_i$ and Alice. $\Box$

\textbf{Remark 1} Since our scheme is an improvement based on the scheme \cite{C32}, it can be said that there are no internal fraudsters in the classical protocol part, that is, the participants are honest. Because if there are dishonest participants showing false shares, they will be recognized by Alice in Step 5.

\begin{flushleft}
And because our scheme is an enhancement of the scheme \cite{C32}, it also has the confidentiality of this scheme, as presented in the following theorem.
\end{flushleft}

\textbf{Theorem 2} (Confidentiality) The proposed scheme satisfies the confidentiality. That is, the external adversary cannot obtain any information about the private keys $s$. And when $h>t(t-1)$, then $t$ or more than $t$ participants with sub-shares can recover the private key $s$, but less than $t$ participants with sub-shares cannot obtain any information about the private key $s$.

\section{The Description of the agreement}

\subsection{Design Issues for System Models and Protocols }
Let $B=\{Bob_1,\cdots,Bob_t\}$. The purpose of this protocol is that Alice wants $t(t\leq n)$ participants from the set $B$ to share a classical key sequence $K$ over a finite field $F_p$.

Our system model is organically composed of a classical network and a quantum network, where the classical network adopts a Shamir $(t,n)$ threshold scheme based on a binary polynomial, the secrets recovered by the participants and the session keys obtained by them are used for the relevant parameters in the quantum network protocol to ensure that it has verifiability of the participants' identities and security of the information.

In the classical network, we assume that each participant is connected to a public authenticated broadcast channel C, so that any message sent through C can be received by other participants. The adversary cannot modify a message sent by an honest participant via C, nor can it prevent an honest participant from receiving a message from C. In our protocol, the recovery of private key $s$ first needs a secure channel which will be established between the distributor Alice and the participant to ensure that Alice can securely distribute sub-shares to the participant. Here we assume that this secure channel is a quantum direct communication channel. Then the channel between any two sub-secret holders must also be secure to ensure the secure exchange of messages, otherwise, other participants can also get recovered secrets. Here the secure channel between every two participants is established by binary polynomial in order to be protected from outside adversary attacks. Thus, in the classical part of our proposed protocol, its security analysis only needs to consider the attacks from the internal participants.

In the quantum system model, we assume that each terminal Bob$_i(i=1,2,\cdots,$ $n)$ also has the ability which can generate, manipulate and measure single photon. For convenience, we assume that this protocol consists of Alice distributing the classical key sequence $K$ to $t$ participants Bob$_1,\cdots$,Bob$_t$ from the set $A$. In the agreement, we agree that t participants must follow the protocol rules and procedures; and when Alice sends a piece of key information to the participants, if the error rate of the received information is lower than a certain threshold. The quantum channel in our agreement can ensure that participants will receive a sub-sequence in this quantum key sequence, that is, it is part of the correct information.

The $d$ group of base used in this protocol is shown in Eq.(2) and Eq.(3), where $|c^{(j)}_l\rangle\in C^p$. From (4), it can be seen that when $y\neq0 \mbox{mod }p$, the base vector of these $p$ groups are transformed into another base vector by applying the unitary transformation $U^y_x=X^xY^y$ to the base vector of these $p$ groups. The relationship between them can be expressed by the superscript or subscript of the base vector via a modulo $p$ operation.

\subsection{Data shared by participants in the classical channel}

\subsubsection{Session keys between two participants}
The distributor Alice chooses an asymmetric binary polynomial $F(x,y)$, see Eq.(5) in Section 2.3 of this paper.

Alice computes $F(x_i,y)$mod $d$ and $F(x,x_i)$ mod $d$ as Bob$_i$'s secret shares, $i\in \{0,1,\cdots,t\}$, $x_i$ is Bob's public identity. Alice sends $F(x_i,y)$ mod $d$ and $F(x,x_i)$ to Bob$_i$ over the classically secure authentication channel. $F(x_i,x_j)$ is used as the encryption key to the encryption function $E$, and $F(x_j,x_i)$ is used as the identification and the encryption key of the quantum states to the data message between Bob$_i$ and Bob$_j$, where $i,j\in \{0,1,\cdots,n\}$ and $i<j$.

\subsubsection{Private keys shared between participants}
The private key to be shared between participants is $s=F(0,0)$, where the recovery of $s$ is described in Section 3.3.
\subsection{Quantum key distribution}
First we assume that Alice has shared the classical data with the participants from the set B=\{Bob$_1,\cdots$, Bob$_t$\}. Suppose that Alice wants to distribute a classical key sequence $K$ consisting of $mp$ elements over a finite field $F_p$ to participants Bob$_1,\cdots$, Bob$_t$ via a quantum channel, where $m=\lfloor p/3\rfloor$.

The proposed multiparty QKA protocol based on duplicate codes \cite{C33} can be described as follows.

\subsubsection{Identification phase}
Alice sends $E_{F(x_0,x_1)}(x_0,t_0,F(x_1,x_0))$ to Bob$_1$ on broadcast channel C, and Bob$_1$ decrypts $E_{F(x_0,x_1)}(x_0,t_0,F(x_1,x_0))$ with $D_{F(x_0,x_1)}$ when he receives it, and he verifies her datas $(x_0,F(x_1,x_0))$. If the datas are correct, Bob$_1$ tells Alice that she has received it, where $t_0$ is the time point on which Alice will send to Bob$_1$ the sequence of quantum states. If $(x_0,F(x_1,x_0))$ is wrong, he will reject this communication.

\subsubsection{Passing random sequences through broadcast channels}
Alice randomly selects a $p$-tuple $(r^{(0)}_1,r^{(0)}_2,\cdots,r^{(0)}_p)$, $i\in \{1,2,\cdots,p\}$, and sends

\begin{equation}
E_{F(x_0,x_1)}(r^{(0)}_1,r^{(0)}_2,\cdots,r^{(0)}_p)
\label{Eq:6}
\end{equation}

\begin{flushleft}
to Bob$_1$ over broadcast channel C, where (6) represents encryption using key $F(x_0,x_1)$ about random sequence $(r^{(0)}_1,r^{(0)}_2,\cdots,r^{(0)}_p)$. When Bob$_1$ receives (6), he will decrypt (6) with $D_{F(x_0,x_1)}$ and tell Alice that it has been received.
\end{flushleft}

\subsubsection{Encoding and distribution of quantum keys}
First the process of Alice distributing a key to Bob$_1$ is given below.

\textbf{Step 1} Alice randomly generates the classical key sequence consisting of $mp$ elements
\begin{equation}
K=(k_1,k_2,\cdots,k_{mp}),
\label{Eq:7}
\end{equation}

\begin{flushleft}
where $k_i\in F_p$, $i=1,2,\cdots,mp$. Then she generates the following sequence from the key sequence $K$.
\end{flushleft}

\begin{equation}
\begin{aligned}
K^{(1)}&=(k^{(1)}_{1,1},k^{(1)}_{1,2},k^{(1)}_{1,3},\cdots,k^{(1)}_{m,1},k^{(1)}_{m,2},k^{(1)}_{m,3}),\\
K^{(2)}&=(k^{(2)}_{1,1},k^{(2)}_{1,2},k^{(2)}_{1,3},\cdots,k^{(2)}_{m,1},k^{(2)}_{m,2},k^{(2)}_{m,3}),\\
&\vdots\\
K^{(p)}&=(k^{(p)}_{1,1},k^{(p)}_{1,2},k^{(p)}_{1,3},\cdots,k^{(p)}_{m,1},k^{(p)}_{m,2},k^{(p)}_{m,3}),\\
\end{aligned}
\label{Eq:8}
\end{equation}

\begin{flushleft}
for $j\in\{1,2,3\}$, where $k^{(t)}_{i,j}=k_{(t-1)m+i}$.
\end{flushleft}

It is easy to see that (8) is an encoding of the information bits in (7) using the ternary repetition code.

\textbf{Step 2} Alice constructs $p$ groups of ordered quantum state sequences consisting of the unbiased bases in (3) according to Eq.(8) as follows:
\begin{equation}
\begin{aligned}
S^{(1)}_A=(|v^{r^{(0)}_1}_{k^{(1)}_{1,1}}\rangle,|v^{r^{(0)}_1}_{k^{(1)}_{1,2}}\rangle,|v^{r^{(0)}_1}_{k^{(1)}_{1,3}}\rangle,&\cdots,|v^{r^{(0)}_1}_{k^{(1)}_{m,1}}\rangle,|v^{r^{(0)}_1}_{k^{(1)}_{m,2}}\rangle,|v^{r^{(0)}_1}_{k^{(1)}_{m,3}}\rangle);\\
S^{(2)}_A=(|v^{r^{(0)}_2}_{k^{(2)}_{1,1}}\rangle,|v^{r^{(0)}_2}_{k^{(2)}_{1,2}}\rangle,|v^{r^{(0)}_2}_{k^{(2)}_{1,3}}\rangle,&\cdots,|v^{r^{(0)}_2}_{k^{(2)}_{m,1}}\rangle,|v^{r^{(0)}_2}_{k^{(2)}_{m,2}}\rangle,|v^{r^{(0)}_2}_{k^{(2)}_{m,3}}\rangle);\\
&\vdots\\
S^{(p)}_A=(|v^{r^{(0)}_p}_{k^{(p)}_{1,1}}\rangle,|v^{r^{(0)}_p}_{k^{(p)}_{1,2}}\rangle,|v^{r^{(0)}_p}_{k^{(p)}_{1,3}}\rangle,&\cdots,|v^{r^{(0)}_p}_{k^{(p)}_{m,1}}\rangle,|v^{r^{(0)}_p}_{k^{(p)}_{m,2}}\rangle,|v^{r^{(0)}_p}_{k^{(p)}_{m,3}}\rangle).\\
\end{aligned}
\label{Eq:9}
\end{equation}

\begin{flushleft}
Alice performs a unitary transformation $U^{r^{(0)}_i}_{k^{(i)}_{1,1}+F(x_1,x_0)}$ on the $i$-th row of the quantum state sequence in (9), and obtains the quantum state sequence
\end{flushleft}

\begin{equation}
\begin{aligned}
\tilde{S}^{(1)}_A=(|v^{r^{(0)}_1}_{k^{(1)}_{1,1}}\rangle,|v^{r^{(0)}_1}_{k^{(1)}_{1,2}}\rangle,|v^{r^{(0)}_1}_{k^{(1)}_{1,3}}\rangle,&\cdots,|v^{r^{(0)}_1}_{k^{(1)}_{m,1}}\rangle,|v^{r^{(0)}_1}_{k^{(1)}_{m,2}}\rangle,|v^{r^{(0)}_1}_{k^{(1)}_{m,3}}\rangle);\\
\tilde{S}^{(2)}_A=(|v^{r^{(0)}_2}_{k^{(2)}_{1,1}}\rangle,|v^{r^{(0)}_2}_{k^{(2)}_{1,2}}\rangle,|v^{r^{(0)}_2}_{k^{(2)}_{1,3}}\rangle,&\cdots,|v^{r^{(0)}_2}_{k^{(2)}_{m,1}}\rangle,|v^{r^{(0)}_2}_{k^{(2)}_{m,2}}\rangle,|v^{r^{(0)}_2}_{k^{(2)}_{m,3}}\rangle);\\
&\vdots\\
\tilde{S}^{(p)}_A=(|v^{r^{(0)}_p}_{k^{(p)}_{1,1}}\rangle,|v^{r^{(0)}_p}_{k^{(p)}_{1,2}}\rangle,|v^{r^{(0)}_p}_{k^{(p)}_{1,3}}\rangle,&\cdots,|v^{r^{(0)}_p}_{k^{(p)}_{m,1}}\rangle,|v^{r^{(0)}_p}_{k^{(p)}_{m,2}}\rangle,|v^{r^{(0)}_p}_{k^{(p)}_{m,3}}\rangle).\\
\end{aligned}
\label{Eq:10}
\end{equation}

\begin{flushleft}
Then Alice sends the quantum state sequence (10) to Bob$_1$ via the quantum channel at $t_0$ moment.
\end{flushleft}

\textbf{Step 3} When Bob$_1$ receives the Eq.(10), Bob$_1$ performs a unitary transformation $(U^0_{F(x_1,x_0)})^{-1}$ on the $i$-th row of quantum state sequences in Eq.(10), $i=1,2,\cdots,p$, then performs group measurements on the obtained $p$-group quantum state sequence, where the quantum state sequences of the $i$-th row are measured with the $i$-th group of unbiased bases $\{|v^{r^{(0)}_i}_l\rangle|l=1,2,\cdots,p\}$ in (3), and the resulting measurements are recorded as
\begin{equation}
\begin{aligned}
L^{(1)}_1=\{l^{(1)}_{1,1},l^{(1)}_{1,2},l^{(1)}_{1,3},&\cdots,l^{(1)}_{m,1},l^{(1)}_{m,2},l^{(1)}_{m,3}\},\\
L^{(2)}_1=\{l^{(2)}_{1,1},l^{(2)}_{1,2},l^{(2)}_{1,3},&\cdots,l^{(2)}_{m,1},l^{(2)}_{m,2},l^{(2)}_{m,3}\},\\
&\vdots\\
L^{(p)}_1=\{l^{(p)}_{1,1},l^{(p)}_{1,2},l^{(p)}_{1,3},&\cdots,l^{(p)}_{m,1},l^{(p)}_{m,2},l^{(p)}_{m,3}\},\\
\end{aligned}
\label{Eq:10}
\end{equation}

\begin{flushleft}
here $l^{(k)}_{i,j}\in F_p$.
\end{flushleft}

\textbf{Step 4} Bob$_1$ counts the data in (11), and lets

$S^{(1)}_k=\{(l^{(k)}_{i,1},l^{(k)}_{i,2},l^{(k)}_{i,3})|$ The three components are not equal with each other\}.

$S^{(2)}_k$= $\{(l^{(k)}_{i,1},l^{(k)}_{i,2},l^{(k)}_{i,3})|l^{(k)}_{i,j_1}$ =$l^{(k)}_{i,j_2}$, and $l^{(k)}_{i,j_1}\neq l^{(k)}_{i,j_3}\}$, where $j_i, j_2, j_3\in \{1,2,3\};$

$S^{(3)}_k=\{(l^{(k)}_{i,1},l^{(k)}_{i,2},l^{(k)}_{i,3})|l^{(k)}_{i,1}=l^{(k)}_{i,2}=l^{(k)}_{i,3}\};$

\begin{flushleft}
Here we define two thresholds $0\leq \varepsilon_i\leq1 (i=1,2)$, and calculate $\frac{|S^{(3)}_1|}{|S^{(1)}_1|+|S^{(2)}_1|+|S^{(3)}_1|}$ and $\frac{|S^{(1)}_1|}{|S^{(1)}_1|+|S^{(2)}_1|+|S^{(3)}_1|}$.
\end{flushleft}

Then Bob$_1$ compares the following data in two cases:

Case (a): $\frac{|S^{(3)}_1|}{|S^{(1)}_1|+|S^{(2)}_1|+|S^{(3)}_1|}<\varepsilon_1$, or $\frac{|S^{(1)}_1|}{|S^{(1)}_1|+|S^{(2)}_1|+|S^{(3)}_1|}>\varepsilon_2$;

Case (b): $\frac{|S^{(3)}_1|}{|S^{(1)}_1|+|S^{(2)}_1|+|S^{(3)}_1|}>\varepsilon_1$, and $\frac{|S^{(1)}_1|}{|S^{(1)}_1|+|S^{(2)}_1|+|S^{(3)}_1|}<\varepsilon_2$.

In case (a), this round will be abandoned.

In case (b), Bob$_1$ will use the error-correction code principle to the set $S^{(3)}_1$ to obtain the following two sets,
\begin{equation}
\begin{aligned}
&S_1=\{(k,i,1)|(l^{(k)}_{i,1},l^{(k)}_{i,2},l^{(k)}_{i,3})\in S^{(3)}_1\},\\
&I_1=\{l^{(k)}_{i,1}|(k,i,1)\in S_1\}.
\end{aligned}
\label{Eq:12}
\end{equation}
Then Bob$_1$ continues next step.

\textbf{Step 5} Through step 4, a sub-sequence $T_1$ of the key sequence $K=(k_1,k_2$, $\cdots,k_{mp})$ can be obtained, where $l^{(t)}_{i,1}=k^{(t)}_{i.1}$, and $k{(t)}_{i,1}$ is from (8). Then Bob$_1$ combines the set $S^{(2)}_1$ to extend the key subsequence $I_1$. This extension is as follows. Let

\begin{eqnarray*}
T_1=\{(k,i,j)|(l^{(k)}_{i,1},l^{(k)}_{i,2},l^{(k)}_{i,3})\in S^{(2)}\},
\end{eqnarray*}

\begin{flushleft}
where $j$ is the least number of 1,2 and 3 such that  $l^{(k)}_{i,j}$ are equal with the two elements of $l^{(k)}_{i,1}$,$l^{(k)}_{i,2}$ and $l^{(k)}_{i,3}$. And let
\end{flushleft}
\begin{eqnarray*}
J_1=\{l^{(k)}_{i,j}|(k,i,j)\in T_1\}.
\end{eqnarray*}

Assuming that the sequence pairs in $T_1$ are already sorted by dictionary order, then take the first sequence pair $(k,i,j)$ of $T_1$ and the corresponding element $l^{(k)}_{i,j}$ of $J_1$ in order and add them to $S_1$ and $I_1$ respectively. Second, Bob$_1$ sends $H_s(x_1,S_1,T_1)$ to Alice, where $s$ is the classical private key recovered by participants. When Alice verifies that this Hash value is correct, she will tell Bob$_1$ this information through the classical channel, and Bob$_1$ will keep this new $S_1$ and $I_1$; otherwise he will restore the previous $S_1$ and $I_1$. The other elements of $T_1$ and $J_1$ are examined in the same way in turn, thus Bob$_1$ will obtain the extended key subsequence $I_1$.

\textbf{Remark 2} Assuming by the quantum model in this paper, Bob$_1$ must be able to obtain a subsequence of the key sequence $K=(k_1,k_2,\cdots,k_{mp})$ using Step 4 when the measured data of (11) meets the conditions of two thresholds. Here this subsequence must be $I_1$ or a proposed subset of $I_1$ according to the error correction theory.

\textbf{Step 6} Alice continues to perform Step 1-5 on the information in $K\backslash I_1$. Finally, she will distributes the key sequence $K=(k_1,k_2,\cdots,k_{mp})$ to Bob$_1$.

\textbf{Step 7} After Bob$_1$ gets the key sequence $K$, Bob$_1$ then does the same process as Alice and passes the key $K=(k_1,k_2,\cdots,k_{mp})$ to $Bob_2$. And so on, eventually $Bob_{t-1}$ passes the key $K$ to $Bob_t$. During the entire pass, the sequence $K$ is recognized by Alice for its correctness. At the same time, the authentication between $Bob_{i-1}$ and Bob$_i$ is guaranteed by the session key $F(x_{i-1},x_i)$ between them and the quantum state transmit moment $t_{i-1}$. For eavesdropping, the test is determined by the error rate $\frac{|S^{(1)}_{i-1}|}{|S^{(1)}_{i-1}|+|S^{(2)}_{i-1}|+|S^{(3)}_{i-1}|}$ derived from the data distribution of the measured value $L^{(1)}_i,L^{(2)}_i,\cdots,L^{(3)}_i$ by Bob$_{i-1}$.

\textbf{Definition 3} The two inequalities in (12) are called the threshold conditions of this protocol.

\textbf{Remark 3} It should be emphasized that although we distribute the key sequence on the finite field $F_p$, the key sequence of 0,1 frequently used such as BB84 protocol can also be represented by the $p$-ary system to correspond to the sequence on $F_p$, thus, can be designed into the above scheme.

\section{Examples}
This section provides a successful example to understand our proposed scheme clearly.
Let the finite field be $F_{11}$. The honest distributor is Alice, and there are 8 shareholders, denoted Bob$_i$, $i=1,2,\cdots,8$. Suppose that Alice wants to share the private key $s=5$ via the classical channel, where Bob$_i$ has a public identity $x_i=i$, and Alice public identity 9. To illustrate the scheme conveniently, we assume that Bob$_1$, Bob$_2$, Bob$_3$ will recover this private key $s$. Similarly, any three participants can restore the private key together in the same way.

Next, we will focus on the process by which Alice sends the key sequence $K$ to Bob$_1$.

\subsection{Data shared by participants in the classic channel}
Alice selects
\begin{align*}
&F(x,y)=7+y+2y^2+y^3+y^4+y^5+y^6+y^7\\
&+2x+xy+xy^2+xy^3+xy^4+xy^5+xy^6+xy^7\\
&+3x^2+x^2y+x^2y^2+x^2y^3+x^2y^4+x^2y^5+x^2y^6+x^2y^7 \mbox{mod }11.
\end{align*}

\begin{flushleft}
and it is obvious that $s=F(0,0)=7$. Then she calculates a pair of sub-shares for herself as follows:
\end{flushleft}

$\begin{cases}
F(x,9)=6+8x+9x^2 \mbox{mod } 11,\\
F(9,x)=4+3x+4y^2+3y^3+3y^4+3y^5+3y^7+3y^7 \mbox{mod } 11,
\end{cases}$

Alice calculates a pair sub-shares for Bob$_1$ as follows:

$\begin{cases}
F(x,1)=4+9x+10x^2 \mbox{ mod } 11,\\
F(1,x)=1+3x+4y^2+3y^3+3y^4+3y^5+3y^7+3y^7 \mbox{ mod } 11,
\end{cases}$
\begin{flushleft}
and sends $F(x,x_1)\mbox{ mod }p$ and $F(x_1,y)\mbox{ mod }p$ to Bob$_1$ through the classically secure authentication channel, and $F(x_0,x_1)$ will be used as the encryption key to the encryption function $E$ between Alice and Bob$_1$, and $F(x_1,x_0)$ is used as identification and encryption key of the quantum states between Alice and Bob$_1$. It is easy to obtain that $F(x_0,x_1)=F(9,1)=4$, $F(x_0,x_1)=F(1,9)=1$.
\end{flushleft}
\subsection{Quantum key distribution}

\subsubsection{Identification phase}
After Bob$_1$ receives $E_4(x_0,t_0,1)$, he decrypts $E_4(x_0,t_0,1)$ with $D_4$. If Bob$_1$ finds that $x_0, F(x_1,x_0)$ is regret, then he will tell Alice that he has received it, where $t_0$ is the time point on which Alice will send the quantum state sequence to Bob$_1$. If Bob$_1$ finds that $x_0, F(x_1,x_0)$ is wrong, he will reject this communication.

\subsubsection{Passing random sequences through broadcast channels}
Alice randomly selects 11 numbers over $F_{11}$ to form a sequence, set it as $(r^{(0)}_{1},r^{(0)}_{2},\cdots,r^{(0)}_{11})=(1,3,6,10,2,1,3,9,6,4,7)$, and sends this random sequence to Bob$_1$ over broadcast channel C using $E_4$. When Bob$_1$ receives this encrypted sequence. He will decrypt it with $D_4$ and tells Alice that it has been received.

\subsubsection{Quantum Key Distribution}
\textbf{Step 1} Alice randomly generates a classical key sequence K, consisting of elements over $F_{11}$, where $K=(k_1,k_2,\cdots,k_{33})$. It can be known that $m=\lfloor\frac{11}{3}\rfloor=3$, so $mp=33$.  Alice generates the following sequence from the key sequence $K$ according to the ternary repetition code.
\begin{equation}
\begin{aligned}
K^{(1)}&=(k^{(1)}_{1,1},k^{(1)}_{1,2},k^{(1)}_{1,3},k^{(1)}_{2,1},k^{(1)}_{2,2},k^{(1)}_{2,3},k^{(1)}_{3,1},k^{(1)}_{3,2},k^{(1)}_{3,3}),\\
K^{(2)}&=(k^{(2)}_{1,1},k^{(2)}_{1,2},k^{(2)}_{1,3},k^{(2)}_{2,1},k^{(2)}_{2,2},k^{(2)}_{2,3},k^{(2)}_{3,1},k^{(2)}_{3,2},k^{(2)}_{3,3}),\\
&\vdots\\
K^{(11)}&=(k^{(11)}_{1,1},k^{(11)}_{1,2},k^{(11)}_{1,3},k^{(11)}_{2,1},k^{(11)}_{2,2},k^{(11)}_{2,3},k^{(11)}_{3,1},k^{(11)}_{3,2},k^{(11)}_{3,3}),\\
\end{aligned}
\label{Eq:13}
\end{equation}

\begin{flushleft}
For $j\in \{1,2,3\}$, where $k^{(t)}_{i,j}=k_{(t-1)m+i}$.
\end{flushleft}
\textbf{Step 2} Alice constructs an ordered sequence of quantum states according to (13).
\begin{footnotesize}
\begin{equation}
\begin{aligned}
S^{(1)}_A=(|v^{r^{(0)}_1}_{k^{(1)}_{1,1}}\rangle,|v^{r^{(0)}_1}_{k^{(1)}_{1,2}}\rangle,|v^{r^{(0)}_1}_{k^{(1)}_{1,3}}\rangle,|v^{r^{(0)}_1}_{k^{(1)}_{2,1}}\rangle,&|v^{r^{(0)}_1}_{k^{(1)}_{2,2}}\rangle,|v^{r^{(0)}_1}_{k^{(1)}_{2,3}}\rangle,|v^{r^{(0)}_1}_{k^{(1)}_{3,1}}\rangle,|v^{r^{(0)}_1}_{k^{(1)}_{3,2}}\rangle,|v^{r^{(0)}_1}_{k^{(1)}_{3,3}}\rangle);\\
S^{(2)}_A=(|v^{r^{(0)}_2}_{k^{(2)}_{1,1}}\rangle,|v^{r^{(0)}_2}_{k^{(2)}_{1,2}}\rangle,|v^{r^{(0)}_2}_{k^{(2)}_{1,3}}\rangle,|v^{r^{(0)}_2}_{k^{(2)}_{2,1}}\rangle,&|v^{r^{(0)}_2}_{k^{(2)}_{2,2}}\rangle,|v^{r^{(0)}_2}_{k^{(2)}_{2,3}}\rangle,|v^{r^{(0)}_2}_{k^{(2)}_{3,1}}\rangle,|v^{r^{(0)}_2}_{k^{(2)}_{3,2}}\rangle,|v^{r^{(0)}_2}_{k^{(2)}_{3,3}}\rangle);\\
&\vdots\\
S^{(11)}_A=(|v^{r^{(0)}_{11}}_{k^{(11)}_{1,1}}\rangle,|v^{r^{(0)}_{11}}_{k^{(11)}_{1,2}}\rangle,|v^{r^{(0)}_{11}}_{k^{(11)}_{1,3}}\rangle,|v^{r^{(0)}_{11}}_{k^{(11)}_{2,1}}\rangle,&|v^{r^{(0)}_{11}}_{k^{(11)}_{2,2}}\rangle,|v^{r^{(0)}_{11}}_{k^{(11)}_{2,3}}\rangle,|v^{r^{(0)}_{11}}_{k^{(11)}_{3,1}}\rangle,|v^{r^{(0)}_{11}}_{k^{(11)}_{3,2}}\rangle,|v^{r^{(0)}_{11}}_{k^{(11)}_{3,3}}\rangle).\\
\end{aligned}
\label{Eq:14}
\end{equation}
\end{footnotesize}

\begin{flushleft}
and does the unitary transformation $U^{r^{(0)}_i}_{k^{(i)}_{1,1}+F(x_1,x_0)}$ on the $i$-th row of the quantum state sequence (14) to obtain the quantum state sequence:
\end{flushleft}
\begin{equation}
\widetilde{S}^{(1)}_A,\widetilde{S}^{(2)}_A,\widetilde{S}^{(3)}_A,\widetilde{S}^{(4)}_A,S^{(5)}_A,\widetilde{S}^{(6)}_A,\widetilde{S}^{(7)}_A,\widetilde{S}^{(8)}_A,\widetilde{S}^{(9)}_A,\widetilde{S}^{(10)}_A,\widetilde{S}^{(11)}_A,
\label{Eq:15}
\end{equation}
\begin{flushleft}
where
\end{flushleft}
\begin{eqnarray*}
\widetilde{S}^{(i)}_A =(&|&v^{r^{(0)}_i}_{k^{(i)}_{1,1}+F(x_1,x_0)}\rangle, |v^{r^{(0)}_i}_{k^{(i)}_{1,2}+F(x_1,x_0)}\rangle, |v^{r^{(0)}_i}_{k^{(i)}_{1,3}+F(x_1,x_0)}\rangle,\\
&|&v^{r^{(0)}_i}_{k^{(i)}_{2,1}+F(x_1,x_0)}\rangle,|v^{r^{(0)}_i}_{k^{(i)}_{2,2}+F(x_1,x_0)}\rangle,|v^{r^{(0)}_i}_{k^{(i)}_{2,3}+F(x_1,x_0)}\rangle\\
&|&v^{r^{(0)}_i}_{k^{(i)}_{3,1}+F(x_1,x_0)}\rangle,|v^{r^{(0)}_i}_{k^{(i)}_{3,2}+F(x_1,x_0)}\rangle,|v^{r^{(0)}_i}_{k^{(i)}_{3,3}+F(x_1,x_0)}\rangle ).
\end{eqnarray*}

\begin{flushleft}
Then Alice sends (15) to Bob$_1$ via the quantum channel at $t_0$ time.
\end{flushleft}

\textbf{Step 3} When Bob$_1$ receives this sequence (15), Bob$_1$ performs a unitary transformation $(U^0_{F(x_1,x_0)})^{-1}$ on the $i$-th row in sequence (15), then performs a group measurement on the sequence (15), e.g., measures $S^{(1)}_A$ using the first measurement basis. And record the measurement result as
\begin{equation}
\begin{aligned}
L^{(r^{(0)}_1)}_1&=(l^{(1)}_{1,1},l^{(1)}_{1,2},l^{(1)}_{1,3},l^{(1)}_{2,1},l^{(1)}_{2,2},l^{(1)}_{2,3},l^{(1)}_{3,1},l^{(1)}_{3,2},l^{(1)}_{3,3}); \\
L^{(r^{(0)}_2)}_1&=(l^{(2)}_{1,1},l^{(2)}_{1,2},l^{(2)}_{1,3},l^{(2)}_{2,1},l^{(2)}_{2,2},l^{(2)}_{2,3},l^{(2)}_{3,1},l^{(2)}_{3,2},l^{(2)}_{3,3}); \\
&\vdots\\
L^{(r^{(0)}_{11})}_1&=(l^{(11)}_{1,1},l^{(11)}_{1,2},l^{(11)}_{1,3},l^{(11)}_{2,1},l^{(11)}_{2,2},l^{(11)}_{2,3},l^{(11)}_{3,1},l^{(11)}_{3,2},l^{(11)}_{3,3}).
\end{aligned}
\label{Eq:16}
\end{equation}

\textbf{Step 4} Bob$_1$ statistics on the data (16). Assuming that the two thresholds are $\varepsilon_1=\frac{1}{2}$ and $\varepsilon_2=\frac{1}{11}$ respectively, and assume that $|S^{(3)}_1|=27$, $|S^{(2)}_1|=3$, $|S^{(1)}_1|=3$, which obtained in this round, then we have
\begin{eqnarray*}
\frac{|S^{(3)}_1|}{|S^{(1)}_1|+|S^{(2)}_1|+|S^{(3)}_1|}=\frac{9}{11}>\frac{1}{2},\\
\frac{|S^{(1)}_1|}{|S^{(1)}_1|+|S^{(2)}_1|+|S^{(3)}_1|}=\frac{1}{11}.
\end{eqnarray*}

\begin{flushleft}
Therefore, the threshold condition of the protocol is satisfied, and thus Bob$_1$ proceeds to the next step.
\end{flushleft}

\textbf{Step 5} For the set $S^{(3)}_1$, Bob$_1$ uses the error-code principle to obtain two sets $S_1$ and $I_1$, we assume that
\begin{footnotesize}
$$ S_1=\left\{
\begin{array}{l}
(1,1,1),(1,2,1),(2,1,1),(2,2,1),(2,3,1),(3,1,1),\\
(3,3,1),(4,1,1),(4,2,1),(4,3,1),(5,1,1),(5,2,1),\\
(6,6,1),(6,2,1),(6,3,1),(7,1,1),(7,3,1),(8,1,1),\\
(8,2,1),(9,1,1),(9,2,1),(9,3,1),(10,1,1),(10,3,1),\\
(11,1,1),(11,2,1),(11,3,1)
\end{array} \right\}, $$
\end{footnotesize}

$$ I_1=\left\{
\begin{array}{l}
l^{(1)}_{1,1},l^{(1)}_{2,1},l^{(1)}_{3,1},l^{(2)}_{1,1},l^{(2)}_{2,1},l^{(2)}_{3,1},l^{(3)}_{1,1},l^{(3)}_{3,1},l^{(4)}_{1,1},l^{(4)}_{2,1},\\
l^{(4)}_{3,1},l^{(5)}_{1,1},l^{(5)}_{2,1},l^{(6)}_{1,1},l^{(6)}_{2,1},l^{(6)}_{3,1},l^{(7)}_{1,1},l^{(7)}_{3,1},l^{(8)}_{1,1},\\
l^{(8)}_{2,1},l^{(9)}_{1,1},l^{(9)}_{2,1},l^{(9)}_{3,1},l^{(10)}_{3,1},l^{(11)}_{1,1},l^{(11)}_{2,1},l^{(11)}_{3,1}\\
\end{array} \right\}. $$

And do that $T_1=\{(1,3,1),(3,2,2),(10,2,2)\}$, $J_1=\{l^{(1)}_{2,1},l^{(3)}_{2,2},l^{(10)}_{2,2}\}$. Bob$_1$ adds (1,3,1) and $l^{(1)}_{3,1}$ to $S_1$ and $I_1$ respectively. Then Bob$_1$ computes $H_s(x_1,S_1,I_1)$ and sends it to Alice. When Alice verifies that the hash value is correct, she will tell Bob$_1$ this information via the classic channel. Bob$_1$ will retain this newly set $S_1$ and $I_1$, thus we have

\begin{footnotesize}
$$ S_1=\left\{
\begin{array}{l}
(1,1,1),(1,2,1),(1,3,1),(2,1,1),(2,2,1),(2,3,1),\\
(3,1,1),(3,3,1),(4,1,1),(4,2,1),(4,3,1),(5,1,1),\\
(5,2,1),(6,1,1),(6,2,1),(6,3,1),(7,1,1),(7,3,1),\\
(8,1,1),(8,2,1),(9,1,1),(9,2,1),(9,3,1),(10,1,1),\\
(10,3,1),(11,1,1),(11,2,1),(11,3,1)
\end{array} \right\}. $$
\end{footnotesize}

\begin{equation}
I_1=\left\{
\begin{array}{l}
l^{(1)}_{1,1},l^{(1)}_{2,1},l^{(1)}_{3,1},l^{(2)}_{1,1},l^{(2)}_{2,1},l^{(2)}_{3,1},l^{(3)}_{1,1},l^{(3)}_{3,1},l^{(4)}_{1,1},l^{(4)}_{2,1},\\
l^{(4)}_{3,1},l^{(5)}_{1,1},l^{(5)}_{2,1},l^{(6)}_{1,1},l^{(6)}_{2,1},l^{(6)}_{3,1},l^{(7)}_{1,1},l^{(7)}_{3,1},l^{(8)}_{1,1},\\
l^{(8)}_{2,1},l^{(9)}_{1,1},l^{(9)}_{2,1},l^{(9)}_{3,1},l^{(10)}_{1,1},l^{(10)}_{3,1},l^{(10)}_{3,1},l^{(11)}_{1,1},l^{(11)}_{2,1},l^{(11)}_{3,1}\\
\end{array} \right\}.
\label{Eq:17}
\end{equation}

Following this method to examine the remaining two elements in $T_1$ and $J_1$ in turn, and finally we can get the extended $S_1$ and $I_1$. Here we assume that the final extended $I_1$ is the $I_1$ in (17).

\textbf{Step 6} Alice continues to perform Step 1-5 on the set $K\backslash I_1$, where it is obvious that $K\backslash I_1=\{k^{(3)}_{2,1},k^{(5)}_{3,1},k^{(7)}_{2,1},k^{(8)}_{3,1},k^{(10)}_{2,1}\}=\{k_{8},k_{15},k_{20},k_{24},k_{29}\}$. Finally she will distribute the key sequence $K=(k_1,k_2,\cdots,k_{33})$ to Bob$_1$.

\textbf{Remark 4} In this example, our classic Shamir threshold scheme is discussed over $F_{11}$, so the dimension of the quantum state space is also 11. In essence, for the formal protocol design, the prime number $p$ can be so large that it can not only ensure the security of the private key $s$ and the session key $F(x_i,x_j)$ between participants, but also can ensure that the classical key sequence $K$ contains more keys for the quantum key distribution.

\section{Security analysis}
Before giving the security analysis, we will give the classical-quantum network diagram of this protocol. See Fig.1.

\begin{figure}[ptb]
\begin{center}
\includegraphics[width=0.5\textwidth]{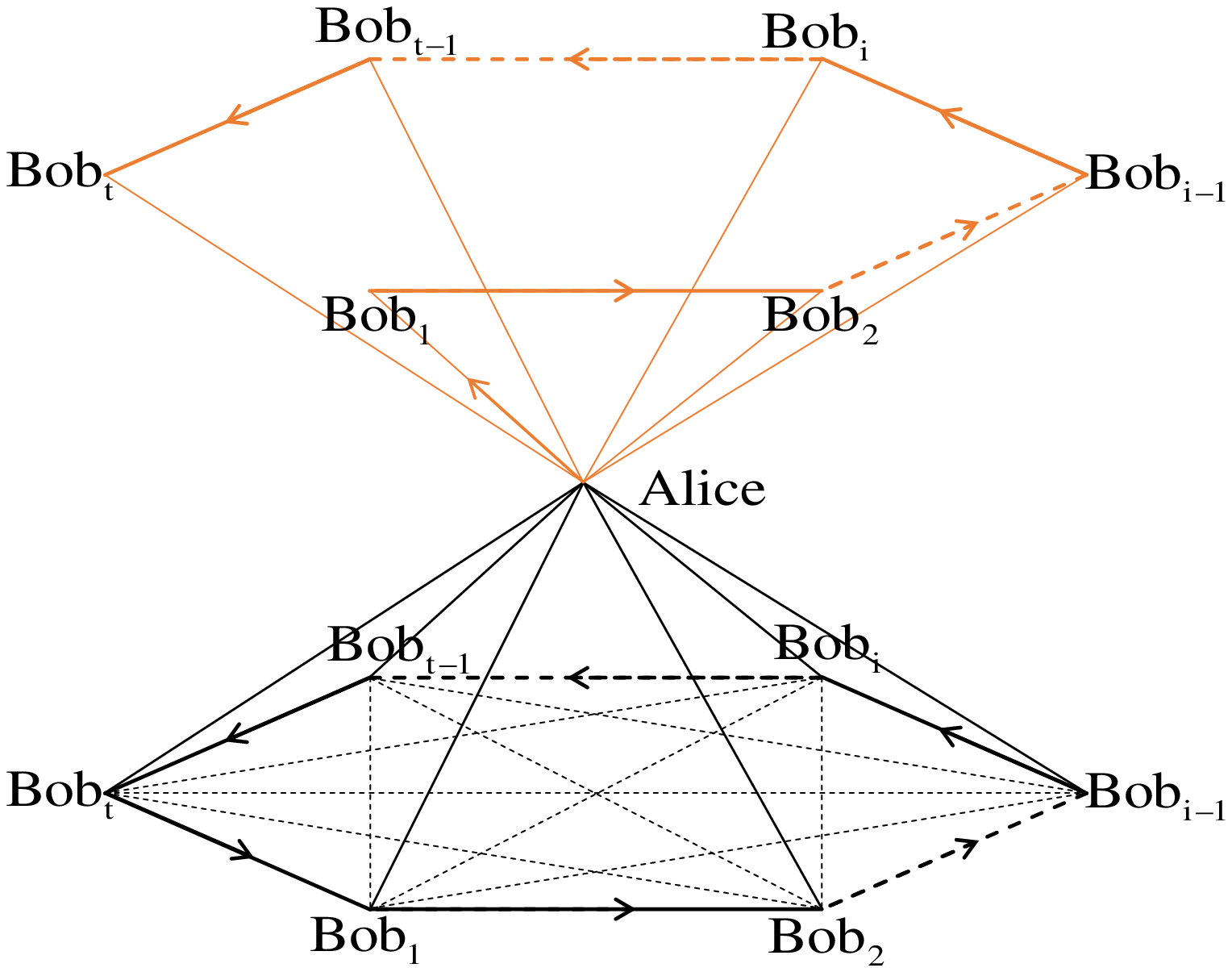}
\end{center}
\caption{The quantum-classical network diagram of this protocal.}
 \end{figure}


\subsection{Correctness verification (classical part and quantum part)}
\textbf{Theorem 3.} The scheme has good accuracy. In other words, in the classic part, the correct private key $s$ can always be reconstructed, and any participant who uses false sub-shares in the sub-share reconstruction stage can be identified. In the quantum part, the participants can always obtain the classical key sequence $K$ eventually if the measured results of the quantum state sequence meet two threshold conditions.

\textbf{Proof} The proof of the correctness in the classical part of this protocol has been shown in Theorem 1. The correctness of the quantum key agreement is given below.

We mainly prove the correctness of the process that Alice sends the classical key sequence $K$ to Bob$_1$ via quantum channel, and the correctness of the process of sending classical key sequence from Bob$_i$ to Bob$_{i+1}$ via quantum channel can be similarly proved.
\begin{equation}
\begin{aligned}
S^{(1)}_A&=(|v^{r^{(0)}_1}_{k^{(1)}_{1,1}+F(x_1,x_0)}\rangle,|v^{r^{(0)}_1}_{k^{(1)}_{1,2}+F(x_1,x_0)}\rangle,|v^{r^{(0)}_1}_{k^{(1)}_{1,3}+F(x_1,x_0)}\rangle,\\
&\cdots,|v^{r^{(0)}_1}_{k^{(1)}_{m,1}+F(x_1,x_0)}\rangle, |v^{r^{(0)}_1}_{k^{(1)}_{m,2}+F(x_1,x_0)}\rangle,|v^{r^{(0)}_1}_{k^{(1)}_{m,3}+F(x_1,x_0)}\rangle);\\
S^{(2)}_A&=(|v^{r^{(0)}_2}_{k^{(2)}_{1,1}+F(x_1,x_0)}\rangle,|v^{r^{(0)}_2}_{k^{(2)}_{1,2}+F(x_1,x_0)}\rangle,|v^{r^{(0)}_2}_{k^{(2)}_{1,3}+F(x_1,x_0)}\rangle,\\
&\cdots,|v^{r^{(0)}_2}_{k^{(2)}_{m,1}+F(x_1,x_0)}\rangle, |v^{r^{(0)}_2}_{k^{(2)}_{m,2}+F(x_1,x_0)}\rangle,|v^{r^{(0)}_2}_{k^{(2)}_{m,3}+F(x_1,x_0)}\rangle);\\
&\vdots\\
S^{(p)}_A&=(|v^{r^{(0)}_p}_{k^{(p)}_{1,1}+F(x_1,x_0)}\rangle,|v^{r^{(0)}_p}_{k^{(p)}_{1,2}+F(x_1,x_0)}\rangle,|v^{r^{(0)}_p}_{k^{(p)}_{1,3}+F(x_1,x_0)}\rangle,\\
&\cdots,|v^{r^{(0)}_p}_{k^{(p)}_{m,1}+F(x_1,x_0)}\rangle,|v^{r^{(0)}_p}_{k^{(p)}_{m,2}+F(x_1,x_0)}\rangle,|v^{r^{(0)}_p}_{k^{(p)}_{m,3}+F(x_1,x_0)}\rangle).\\
\end{aligned}
\label{Eq:10}
\end{equation}

From the process that the key sequence $K=(k_1,k_2,\cdots,k_{mp})$ is encoded, and then sent to Bob$_1$, We can know that the quantum state sequence $\widetilde{S}^{(1)}_A,\widetilde{S}^{(2)}_A,\cdots,\widetilde{S}^{(p)}_A$ from (10) are sent to Bob$_1$ via quantum channel. It's obvious that the classical key sequence $K=(k_1,k_2,\cdots,k_{mp})$ is hidden in each subscript of the unbiased bases, and it is encrypted with $F(x_1,x_0)$ at the same time, the superscripts are encrypted with random sequence $(r^{(1)}_1,r^{(2)}_1,\cdots,r^{(p)}_1)$ respectively, where $F(x_1,x_0)$, and $(r^{(0)}_1,r^{(0)}_2,\cdots,r^{(0)}_p)$ are sent via secure channel. Therefore, it is impossible that $Bob_i (i\neq1)$ deciphers the key sequence information by intercepting these particles. That is to say, after receiving (10), only Bob$_1$ can implement the correct unitary transformation and measurement, and obtain the information related to the key sequence. When the data obtained meet the two threshold conditions in Definition 3, we assume that the subsequence obtained from the key sequence $K$ satisfies this protocol model, then $S_1=\{(k,i,1)|(l^{(k)}_{i,1},l^{(k)}_{i,2},l^{(k)}_{i,3})\in S^{(3)}_1\}$ is not empty. This is because that each triplex code in the set must contain a certain information of $K=(k_1,k_2,\cdots,k_{mp})$ using the error correction principle of triplex code; For each ordered group in $S^{(2)}_1=\{(l^{(k)}_{i,1},l^{(k)}_{i,2},l^{(k)}_{i,3})|$ There are two and only two elements which are equal$\}$. Essentially, there is only one error which can be corrected according to the error correction principle of the triple code, that is, the error bit is consistent with the other two bits. However, to further ensure the accuracy of this information, we need to certify Alice when an extended subset $I_1$ of elements from the set $S^{(2)}_1$ is obtained, thus this extension $I_1$ is exactly the correct subsequence of the key sequence. Then, do the same process to the sequence as above.

Thus, Bob$_1$ can finally obtain the correct classical key sequence $K=(k_1,k_2,$ $\cdots,k_{mp})$. $\Box$

There might be various attacks by external opponents. However we show that none of theses attacks can get any information about the classical sequence $k$.

\textbf{Theorem 4} (Confidentiality) The scheme meets the requirement of confidentiality. In other words, in the classical protocol part, the external adversary cannot obtain any information about the private key and the session key between participants. In the quantum protocol part, the classical key sequence $K$ cannot be obtained by the external adversary.

\textbf{Proof} This protocol consists of two parts. In the classical protocol part, it can be known from Theorem 3 that the external attacker Eve cannot obtain any information about the private key $s$ from the participants and the session key $F(x,y)$ between Bob$_i$ and $Bob_j$, where $i,j\in \{1,2,\cdots,t\},i\neq j$.

In the quantum protocol part, we discuss the following two cases:

\textbf{1)}	The value of $F(x_1,x_0)$ is not correct, or arrival time for the quantum state sequence is not reasonable. Bob$_1$ can verify Alice's identity according to the values $F(x_1,x_0)$ which Alice sends to him. The value $t_0$ can be used to detect whether or not Eve is eavesdropping. So if the value $F(x_1,x_0)$ is inaccurate or the quantum state sequence doesn't arrive during a reasonable time period, then Bob$_1$ will give up the measure of the quantum states in this round. Therefore, this quantum states sequence $S^{(1)}_A,S^{(2)}_A,\cdots,S^{(p)}_A$ encrypted by a random sequence $(r^{(0)}_1,r^{(0)}_2,\cdots,r^{(0)}_p)$ will be abandoned. For Eve, he might want to get relevant information about the encrypted (10). Next, we will prove that even if Eve made some measurement or other interference on the quantum state sequence from (10), he would not get any information about the classical key sequence.

At this point, Eve may have taken the following attacks:

\textbf{a) Intercept resend attack}

Eve may have intercepted the message particles in the process that Bob$_i$ transfered these particles to Bob$_{i+1}$, and then reemitted their own forged particles to Bob$_{i+1}$, where $i\in\{0,1,\cdots,t-1\}$. First, we calculate the probability that Eve intercepts a message particle and gets the key message successfully. It is known that Eve, the eavesdropper, does not know any information about the measurement base, because this measurement base is sent to each participant through quantum security direct communication in this scheme, and is not disclosed to the public, Eve has to choose one of the relevant measurement bases in order to obtain the original secret. We know that only when the chosen basis is the real measurement basis, she can get the measurement result, which means that the probability of her successfully stealing the measurement basis is $1/p$. At the same time, even if Eve had chosen the right basis, he would have had to perform a correct unitary transformation on the measured particles, but he can only infers the unitary transformation in terms of probabilities $1/p$ . Therefore, the probability that he can intercept a particle and get the correct key information is at most $\frac{3}{p^2}$.

According to the above analysis, the probability of Eve successfully obtaining the key sequence is $(\frac{3}{p^2})^{3pm}$. When $p$ is a large prime numbers, the probability of Eve successful stealing the key information $K$ will becomes very small.

\textbf{b) Entanglement measurement attack}
Eavesdropper Eve entangled the auxiliary quantum state to the transmitted quantum state, or replaced the quantum state with a new entangled state. However, the entanglement switching causes these quantum states to be indistinguishable, he could not get any information of the key, and the entanglement measurement attack was invalid.

Therefore, even if Eve makes measurements or other disturbances on this round of quantum state sequence, he will not get any information about the classical key sequence K. Moreover, in the next round of quantum sequence transfer, Alice will re-select the random sequence $(r^{(0)}_1,r^{(0)}_2,\cdots,r^{(0)}_p)$ to encode and encrypt the classical sequence K, so that Eve will perform this round of quantum state sequence from (10). So the information obtained will not help him in the next round of measurement.

\textbf{2)} If the value of $F(x_1,x_0)$ is correct and the arrival time of the quantum state sequence is reasonable, it can be divided into the following two cases according to the measurement results of the quantum state (10):

\textbf{2.1)} The measurement results does not satisfy the threshold condition in Definition 3. In this case, Bob$_1$ will still discard the quantum state sequence from (10) delivered in this round.

\textbf{2.2)} The measurement results satisfy the threshold condition in Definition 3. At this time, Bob$_1$ will get the subsequence $I_1$ of the classical sequence $K$ from this round of measurement.

In both cases 2.1) and 2.2), even if Eve does some measurements or other interference with this round of quantum state sequence, since the quantum state sequence is encrypted, according to the two attacks taken by Eve in 1), it can also prove that Eve has no information about the classic key sequence $K$ in this round.

By the above analysis, it can be shown that Eve cannot obtain any information of the classic key sequence $K$ during the process of Alice passing this key sequence to Bob$_1$. The same is true for the process of passing the key sequence $K$ from Bob$_i$ to $Bob_{i+1}$ about its confidentiality. Thus Eve's information obtained on the quantum state (10) in this round will not be of any help to him in the next round of measurement. $\Box$

\section{Security Comparison}

\subsection{Efficiency}
Efficiency is an important indicator of the agreement. Table 1 shows the comparison of several multi-party QKA agreements with ours. Note that the efficiency values in Table 1 are calculated according to the definition of Cabello efficiency \cite{A25}. Since one of our main contributions is to improve efficiency and reduce quantum resource consumption, we focus on comparing with the schemes in references \cite{A25,A26,A27,A28}. It should be noted that we assume that the number of participants is $t$ .

The Cabello efficiency of the QKA protocol is defined as $\eta=c/{a+b}$, where $c$, $a$ and $b$ represent the number of shared classical bits, the number of qubits used, and the classical exchange number of bits (except monitoring). In our protocol, the efficiency is $\eta={mp}/{3mpt}=1/{3t}$, where the number of bits of the shared key is $mp$, and the total number of qubits used by $t$ participants is $3mpt$.

\begin{table}[h]
\centering
\caption{Comparison about the efficiency among several kinds of multi-party QKA protocols.}
\label{tab:1}
\renewcommand\tabcolsep{1.2pt}
\scriptsize
\begin{tabular}{|l|c |c |c |c|}
\hline
 & OR & Quantum communication & NQO & QE \\\hline
Ref. [25] & GHZ states     & One-way            & SQM         & $1/{2t(t-1)}$ \\
Ref. [26] & Single photons & Two-way            & FQOM + SQUO & $1/{2t(t-1)}$ \\
Ref. [27] & Cluster states & $\frac{t(t+1)}{2}$ & FQOM + SQUO+ CBM  & $1/2t$ \\
Ref. [28] & Bell states    & One-way          & BSM        & $1/{(3\times2^{t-1})}$ \\
Ours      & Single photons & t-1              & SQM + SQUO  & $1/{3t}$ \\\hline
\end{tabular}
\begin{center}
        \footnotesize
        QR quantum resource, NQO necessary quantum operation, QE quantum efficiency, SQUO single-qubit unitary operation, SQM single-qubit measurement, FQOM four-qubit orthogonal measurement, BSM Bell-basis measurement, CBM cluster basis measurement.
      \end{center}
\end{table}

As shown in Table 1, only the scheme in reference \cite{A26} is more efficient with us than our agreement. However, the solution in reference \cite{A26} requires more quantum resource costs than the QKA protocol we proposed here. First, the information of the protocol is carried by the Cluster entangled state, and the $t$ participants are involved in ${ t(t+1)}/2$ quantum communication, and complex cluster orthogonal ground state measurement. As we all know, quantum resources are more expensive than classical resources. Therefore, our scheme is more economical than the scheme in reference \cite{A26}.

\subsection{Verifiability}
The following Table 2 is consistent with the references cited in Table 1, and mainly shows the detailed comparison between our protocol and the scheme in references \cite{A25,A26,A27,A28} in terms of participant authentication, information transmission encryption, and digital authentication. Participant's identity authentication means that the recipient can judge the identity of the sender based on the obtained quantum state information; the transmission encryption of information means that the quantum state sent by the sender is encrypted through some unitary transformation and other technologies; digital authentication means that the key information finally obtained by the participants is obtained through encryption.

\begin{table}[h]
\caption{Comparisons about the verifiablity among several kinds of multi-party QKA protocols.}
\label{tab:2}
\centering
\renewcommand\tabcolsep{1pt}
\scriptsize
\begin{tabular}{|l|c|c|c|}
\hline
 & Identity authentication & Digital authentication & Transmission encryption \\\hline
Ref. \cite{A25} & YES    & YES   & YES \\
Ref. \cite{A26} & YES    & YES   & YES \\
Ref. \cite{A27} & NO     & NO    & NO \\
Ref. \cite{A28} & YES    & NO    & NO \\
Ours     & YES    & YES   & YES \\\hline
\end{tabular}
\end{table}

Although some quantum states are unlikely to be attacked by external participants during the transmission process, after we encrypt them, the probability of such an attack will become even smaller, and our solution is based on the encryption of these quantum states. During the transfer process, the identity of the participant was also verified. Therefore, our protocol is more secure than the above-mentioned protocols.
From the analysis of Table 1 and Table 2, we can see that our protocol has more advantages in terms of efficiency, quantum resources and security by comparing with the literature \cite{A25,A26,A27,A28}.

\section{Conclusion}
A multi-party quantum key distribution protocol based on repetitive code is designed in this paper. The classical key sequence is encoded by repetitive code, the corresponding quantum state sequence is unitary transformed with the parameters from this threshold protocol, and these unitary transformations make the sequence encrypted. Although the information of this protocol are some elements over the finite field with odd prime numbers, it can also be used for the transmission of information sequences consisting of 0 and 1. First, the information consist of 0 and 1 can be transformed into some element over a finite field , then it can also be converted into a corresponding sequence of 0's and 1's when this information is performed through our proposal.

Compared with some existing protocols, our protocol has the function of verifying the identity of participants, and it does not need decoy state particles in the detection of eavesdroppers. Thus, the efficiency of our protocol is obviously improved, the quantum resources are also saved, and the classical scheme used in this scheme has a lower computational complexity. This protocol provides a general and practical quantum multi-party quantum key distribution scheme, which will be expected to be widely used in the future quantum communication environment.
\\
\\

\end{document}